\documentclass{elsart5p}

\usepackage{graphicx}
\usepackage{amssymb}

\usepackage{times}  

\begin{document}

\begin{frontmatter}



\title{Dependence of band structures on stacking and field in layered graphene}


\author{Masato Aoki}
%\author{Masato Aoki\corauthref{cor1}}
\ead{masato@gifu-u.ac.jp}
\author{and Hiroshi Amawashi}
\address{Faculty of Engineering, Gifu University
Yanagido, Gifu 501-1193, Japan}
%%%%\corauth[cor1]{Corresponding author.}
\begin{abstract}
% Text of abstract
 
Novel systems of layered graphene are attracting interest 
for theories and applications. The stability, band structures of 
few-layer graphite films, and  their 
dependence on electric field applied along the c-axis are examined
 within the density functional theory.
We predict that those of Bernal type and also rhombohedral type
tri- and tetra-layer graphite films exhibit stability. 
The rhombohedral-type systems including
AB-bilayer, show variable band gap induced
by perpendicular electric field, whereas the other
systems such as the Bernal-type films stay semi-metallic.

\end{abstract}

\end{frontmatter}

% main text
\section{Introduction}
\label{sec:Introduction}

%% Background : 
%%  few-layer graphite is sparking interest
%%  Which points are of interest?
Amongst the variety of allotropes of carbon over the full range of
dimensionality, two-dimensional (2D) crystals consist of a single or few
graphite layers have been stimulating scientist's interest in 
theories and in possibility for novel applications.
The observations of anomalous electric field effect on 
Hall coefficient\cite{Novoselov-Sci04}, unconventional integer
quantum Hall effects in monolayer\cite{Zhang-nat05} and 
bilayer graphite\cite{Novoselov-NP06} have been enabled by
recent achievement of high quality preparation 
of the ultra-thin graphitic films\cite{Novoselov-Sci04}.

%% band structure has fundamental importance
%% existing results by some groups
These unusual electronic properties are believed to originate
in unusual band structures near Fermi level of graphene\cite{Zheng-Ando-PRB02}, 
a graphitic monolayer, and of its layered structures.
Graphene has a direct zero energy gap between $k$-linear 
$\pi$ bands crossing at each corner (K point) of 
the 2D hexagonal Brillouin zone.  The charge carriers in
graphene therefore are massless Dirac fermions, which leads 
to shifted Hall plateaus\cite{Zhang-nat05}. While the chiral 
parabolic band dispersion in AB-stacked bilayer graphene was 
found to give another new type of Hall effect\cite{Novoselov-NP06}.

Very recently
several groups have reported calculated low-energy
band structures of AB-stacked graphene layers using 
Tight-Binding (TB) models for 
$\pi$ band at zero electric field\cite{Partoens-PRB06}
and with field perpendicular to the layer\cite{Lu-JPCM06}.
In the latter, they examined field induced band gap in AB-bilayer
and found interesting band deformation in ABA- and ABAB-stacked
systems.

However, reliable TB parameters are only available
for the limited equilibrium structures. 
Furthermore, the self-consistency should be crucial
for studying their response against the electric field, 
because of the induced charge redistributions or polarizations
as pointed out in the studies of dielectric function 
of graphite\cite{Marinopoulos-PRB04}. 
Properties  of non-AB-type systems
must be worth examining, because 16 percent
of the natural and synthesized graphite are
rhombohedral (ABC-stacked) with other 4 percent
disordered.

The purpose of the present study is to examine the band
structures and interlayer binding energy of 
few-layer graphite (FLG) 
within the density functional theory (DFT) with  
the local density approximation (LDA), 
focusing on the effects of stacking sequence and external 
electric field.

In the next section, we summarize the method of calculation,
then possible equilibrium structures are sought in Sec.~\ref{sec:Structures}.
Section~\ref{sec:band} illustrates the calculated band structures
near the Fermi level, followed by a concluding section.

\section{Methods}
\label{sec:Method}

%%
%% Computer code, method, approximations used
%%
We  employed the CPMD pseudopotential planewave code\cite{CPMD}
as the {\it ab initio} total energy and band calculations 
in the present study. 
Goedecker-Teter-Hutter pseudopotential for carbon together with 
Pad\'e approximant LDA exchange-correlation potential
were used\cite{Goedecker-PP-PRB96}.
A sufficiently large planewave cutoff of 120 Ry was used to achieve
convergence for the wavefunctions, and 480 Ry for the density.
The Monkhorst-Pack\cite{Monkhorst-Pack-PRB76} 
special $k$-point mesh of $10\times10\times1$ 
within the first Brillouin zone (BZ) was used in the density 
optimizations.

\begin{figure}[tb]
\begin{center}
\includegraphics[width=60mm, trim=0 0 0 0]{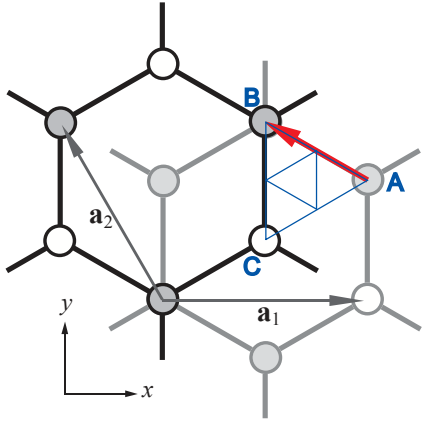}
\end{center}
\caption{An in-plane translation ($\mathrm{A to B}$) of an upper 
graphene layer relative to the bottom layer. Triangle ABC
defines the periphery of irreducible translations.}
\label{displace2}
\end{figure}

Two-dimensional (2D) structures of two up to four graphene layers 
are simulated using 3D hexagonal unit cells defined by primitive
lattice vectors (see Fig.~\ref{displace2})
$\mathbf{a}_1=a(1,0,0)$, $\mathbf{a}_2=a(-\frac{1}{2},\frac{1}{2}\sqrt{3},0)$, 
and $\mathbf{a}_3=(0,0,c)$ 
with the optimized parameter $a=2.4595$ {\AA}. 
and $c=$ $15$, $20$ and $25$ {\AA} for bi-, tri- and tetra-layer systems, respectively,
which ensure at least 10 {\AA} of vacuum separations along c-axis.

The primitive reciprocal vectors in the basal plane are
$\mathbf{b}_1=(2\pi/a)(1,1/\sqrt{3})$, 
$\mathbf{b}_2=(2\pi/a)(0,2/\sqrt{3})$,
 and the corners of the first Brillouin zone are labeled ``K'' which
are given by $\frac{1}{3}(\mathbf{b}_1+\mathbf{b}_2)$ 
plus the arbitrary reciprocal vectors.

We examine the  stability of FLG systems using the 
total energy calculation within the DFT-LDA. It is known
that DFT-LDA fails to describe long-range dispersion (van der Waals)
interactions, but it well describes
the short-range repulsion \cite{Girifalco-PRB02} and
predicts the ground state structure 
and interlayer binding energy \cite{Schabel-Gr-PRB92}
in good agreement with experiments.
Furthermore, Lennard-Jones type van der Waals energy  
does not make much difference between a discreat-atom and 
continuum models for graphitic plane. 
The zero-point phonon energy 
may not be important in the present calculations,
since the influence of stacking configuration on it
is considered to be negligible.

\section{Structures of FLGs}
\label{sec:Structures}

\begin{figure}[tb]
\begin{center}
\includegraphics[width=88mm, trim=0 0 0 0]{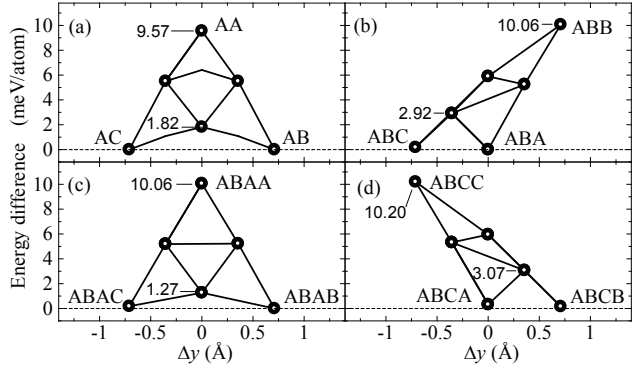}
\end{center}
\caption{Energy surface of FLGs, 
(a) AX-bilayer, (b) ABX-trilayer, (c) ABAX-tetralayer and
(d) ABCX-tetralayer,
as a function of
in-plane displacement of the top graphene overlayer (X).
The total energies of the Bernal-type (AB, ABA and ABAB) stacking 
are taken as reference energies. 
Note that energies are represented in meV per carbon atom on 
the top layer.
} 
\label{stacking}
\end{figure}

Figure~\ref{stacking}(a) shows the calculated total energies
of AX-bilayer graphite with different stacking configurations 
compared to that of the Bernal AB-bilayer, in which the translation
of the top layer X is B (and equivalently C). The interlayer 
separation (to be referred to as $d$) has been optimized for each structures. 
Note that energies are represented in meV per carbon atom on the top overlayer,
which in fact correspond to the exfoliation energies\cite{Schabel-Gr-PRB92}.
We see that AA structure marks, as is well-known, the highest 
energy of about 10 meV/atom with $d=3.62$ {\AA} compared to the 
lowest energy of AB (AC) structure with $d=3.33$ {\AA}.
This energy of AA-bilayer is as large as a half of the 
exfoliation energy observed (22.8 meV/atom) or predicted (24 meV/atom) 
for graphite\cite{Schabel-Gr-PRB92}. A translation of the top layer X
from the in-plane position B to the nearest minimum C will 
encounter adiabatic energy barrier of 1.82 meV/atom at the midpoint, 
which is about 20 percent of the energy increase at 
the AA stacking.

This kind of barrier height is more pronounced (2.92 meV/atom) 
in ABX-trilayer graphite as shown in Fig.~\ref{stacking}(b), 
in which the structure of the first 
two graphene layers are fixed at the AB-bilayer optimized by itself and
the top layer is allowed to translate. The separations from the substrate
AB-bilayer are optimized. ABC-trilayer with optimum separation of 
$d=3.33$ {\AA} is found to have a slightly higher energy (0.18 meV/atom) 
than ABA with $d=3.33$ {\AA}.  

The behavior of barrier along the translational paths 
(Figs.~\ref{stacking}(c) and (d)) is
interesting in tetralayer graphite. The path linking
ABAC (0.18 meV/atom) and ABAB (0 meV/atom by definition) 
has only a reduced barrier peak of 1.27 meV/atom, which
would make transformation to Bernal stacking easier, 
whereas the path from ABCA (0.33 meV/atom) to ABCB (0.18 meV/atom) 
has a much harder barrier peak of 3.07 meV/atom. Since ABCB is 
equivalent to ABAC, we find that the most stable 
Bernal-type ABAB-tetralayer and the second most stable 
rhombohedral-type ABCA-tetralayer is separated by a translational
barrier of about 3 meV/atom, if only a surface layer is allowed to
move. These stable structures have a common optimum interlayer separation
of $d=3.33$ {\AA}.
This result is consistent with the observation of rhombohedral
structure in natural and synthesized graphite\cite{Schabel-Gr-PRB92},
and possibility of fabricating non-Bernal tri- and tetra-layer graphene 
films can be speculated.
We have examined the effect of external electronic field on
the relaxed structures of these FLGs. However, no discernible change
in the order of stability and interlayer separations was found.

\section{Band structures and field effect}
\label{sec:band}

For FLGs that have been found globally or locally minimum in total energy
({\it i.e.}, AB-bilayer, ABA- and ABC-trilayer, and ABAB-, ABCA-, ABAC-, 
ABCB-tetralayer systems), 
we have calculated the band structures of the systems in the zero field
and also in an uniform external electric field of strength 
$\mathcal{E}=77.94$ mV/{\AA}, applied perpendicular to the graphene plane.
This field was represented by an additional sawtooth potential, which is
 piecewise
proportional to $z$ with discontinuity at the mid-plane of the vacuum region
in the cell.
The potential difference applied between two adjacent graphene planes
is $\Delta{V}=0.2595$ Volts.

\subsection{AB-bilayer graphite}

\begin{figure}[tb]
\begin{center}
\includegraphics[width=88mm, trim=0 0 0 0]{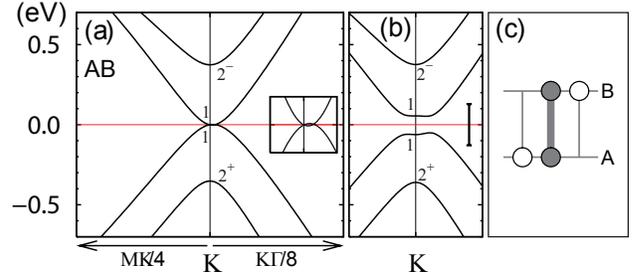}
\end{center}
\caption{Band structures of AB-bilayer graphite film: 
(a) without field, (b) with field of $\mathcal{E}=77.94$ mV/{\AA}, 
(c) shows TB diagrams at the K point.
The inset magnifies 1/8 range of (a) about K and energy range of $\pm30$ meV.
For indices 1 and $2^\pm$ see text.
}
\label{band-AB}
\end{figure}

The band structure near the corner of BZ (K point) 
of AB-bilayer graphite with no external field is semi-metallic
with chiral parabolic dispersions as shown in Fig.~\ref{band-AB}(a). 
The effect of perpendicular field $\mathcal{E}$ 
 opens an energy gap as shown in Fig.~\ref{band-AB}(b). 
The gap at the K point may indeed be 
understood in terms of the TB
description  (such as the SWMcC model\cite{CharlierMG92}) 
of $\pi$ electrons as a lifted degeneracy of non-bonding
$p_z$ orbitals, as a consequence of the effective field 
that breaks the symmetry along the $c$-axis.
The TB Hamiltonian of AB-bilayer graphite at the K point 
is identical to that for the system consists of one dimer and 
two equivalently isolated atoms as depicted in 
Fig.~\ref{band-AB}(c). 
Indices 1 and $2^{+(-)}$ in Figs.~\ref{band-AB}(a) and (b)
label monomeric non-bonding and dimeric bonding(antibonding) states at the K point. 
The hopping parameter for the dimer, 
which can be determined by a half
the difference in eigenvalues of $2^{\pm}$ states at zero field, is found
to be $\gamma_1=0.363$ eV in agreement with the well-known value of
the SWMcC model.

We point out that the gap at the K point, 
$E_{g}^{\mathrm{AB}}(\mathrm{K})=0.117$ eV, is only
45 percent of the unscreened potential difference $\Delta{V}$ 
(indicated by the length of a bar in Fig.~\ref{band-AB}(c))
that the non-self-consistent TB model should predict.
Our calculated result of the gap directly measures the
effective field felt by the AB-bilayer film, and hence
the dielectric response of bilayer graphite film against 
uniform and static perpendicular field, giving the dielectric
constant of $\epsilon(\mathrm{AB})=2.2$. This value of the internal 
effective field was also confirmed from the calculated 
lowest two eigenvalues of $\sigma$ band 
at the BZ center($\Gamma$ point).

In addition, no discernible  influence of the field was found on the
hopping parameter $\gamma_1$, because the field induced gap 
$E_{g}^{\mathrm{AB}}(\mathrm{K})$ was perfectly consistent with
that predicted by $\gamma_1$ at zero field 
and effective potential difference $\epsilon^{-1}\Delta{V}$, 
{\it i.e.}, $2[\gamma_1^2+(\epsilon^{-1}\Delta{V})^2/4]^{1/2}$.
Moreover, this fact was further confirmed in a separate calculation,
in which the strength of the external field was doubled
to widen the band gap at the K point to $0.221$ eV.
Meanwhile, Ohta {\it et al.}\cite{Ohta-Sci06} have reported 
significant change in the dimeric interaction with 
electron doping by potassium adsorption on a surface of AB-bilayer,
from their angle-resolved photoemission experiments with analysis using
TB fitting. This apparent contradiction with the present results
might be related to the shift of Fermi level (electron number) 
due to doping in their system.

\begin{figure}[thb]
\begin{center}
\includegraphics[width=88mm, trim=0 0 0 0]{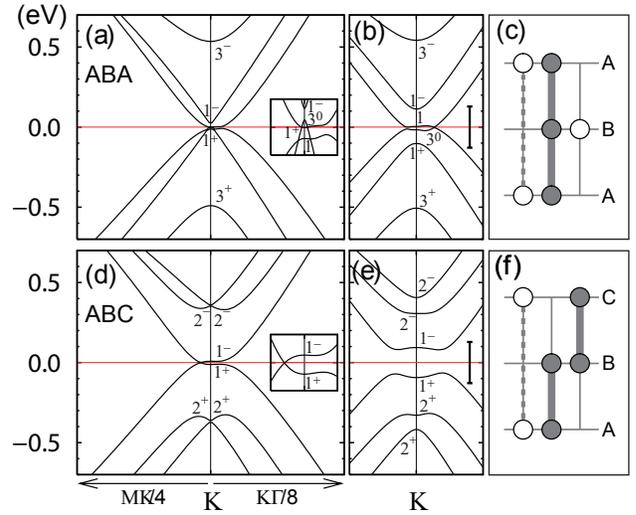}
\end{center}
\caption{Band structures of trilayer graphite films:
(a) ABA-stacked without field, (b) with field of $\mathcal{E}=77.94$ mV/{\AA}, 
(d) ABC-stacked without, and (e) with the field, 
(c) and (f) are TB diagrams at the K point.
The inset in (a) and (d) magnifies 1/8 range about the
K and energy range of $\pm30$ meV.
}
\label{band-3L}
\end{figure}
\begin{figure}[tb]
\begin{center}
\includegraphics[width=88mm, trim=0 0 0 0]{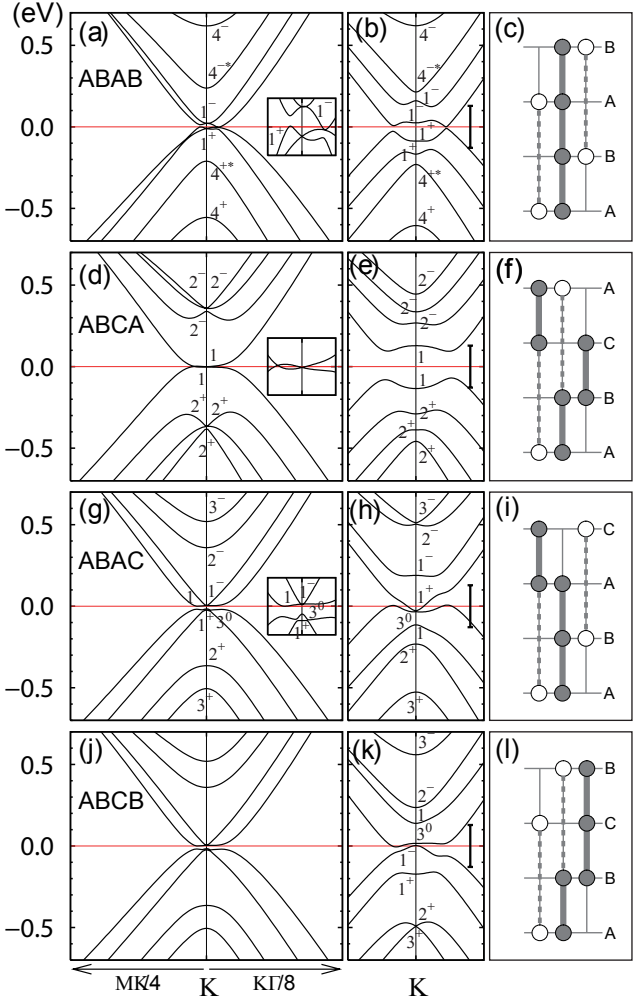}
\end{center}
\caption{Band structures of tetralayer graphite films.
(a) ABAB-stacked without field, (b) with field of $\mathcal{E}=77.94$ mV/{\AA}, 
(d) ABCA-stacked without, and (e) with the field, 
(g) ABAC-stacked without, and (h) with the field, 
(j) ABCB-stacked without, and (k) with the field. 
(c), (f) and (l) are TB diagrams at K point.}
\label{band-4L}
\end{figure}

\subsection{Trilayer graphite}

Figure~\ref{band-3L}(a) shows that
the band structure near the corner of BZ (K point) 
of ABA-trilayer graphite film at zero field
is semi-metallic with rather small overlap of
conduction and valence $\pi$ bands.
The TB Hamiltonian of the ABA-trilayer at the K point is equivalent to
that for a linear trimer(see panel (c)) with hopping for the 
nearest neighbors $\gamma_1$
and for the next neighbor $\gamma_5$ that links both ends, 
and a perfectly non-bonding monomer, and
a very weakly bonded (with $\gamma_2$; a dashed link in panel (c)) 
dimer that opens a tiny gap (12 meV) between nearly $k$-linear branches. 
This trimer at zero field has three states, 
which are to be referred to as bonding/antibonding and non-bonding 
states labeled by $3^{+/-}$ and $3^0$,
with eigenvalues $(\frac{1}{2}\gamma_5 \mp \sqrt{2\gamma_1^2+ \gamma_5^2/4})$
and $-\gamma_5$, respectively. Note that $3^0$ actually is the bonding
state of orbitals at both ends via the second neighbor hopping $\gamma_5$.

On the other hand, ABC-trilayer graphite film at zero field
is as shown in Fig.~\ref{band-3L}(d) a 
zero-gap semiconductor(or zero band-overlap semi-metal) with
a band contact near the K point on the KM symmetry line. 
The gap at the K point is 20 meV, which arise from the splitting
between bonding($1^+$) and antibonding ($1^-$) states
the weakly bonded dimer with the hopping parameter of 
$\gamma_2=10$ meV. The strongly bonded dimers (Fig.~\ref{band-3L}(f))
are giving degenerate states.

The influence of the electric field on band structures of trilayer systems
are shown in Fig.~\ref{band-3L}(d) and (e).
We see the ABA-trilayer stays semi-metallic, because of
the presence of almost degenerate states of non-bonding monomer in the 
middle plane (B) and trimer, which are  nearly independent of the applied field.
Whereas, the ABC-trilayer opens a energy gap 
at the K point, $E_{g}^{\mathrm{ABC}}(\mathrm{K})=0.188$ eV,
and a minimum direct gap 0.146 eV. 
This energy gap is the $1^+$--$1^-$ 
splitting and is nearly double the width
of AB-bilayer's induced by the same external field, reflecting
the doubled thickness of the film, since the $\gamma_2$ is 
order of magnitude smaller than the external potential difference.
Therefore, the ABC-trilayer can be a tunable narrow-gap semiconductor
if fabricated.
The dielectric constant of ABA and ABC trilayers at the present applied field
may be evaluated from  the $1^+$--$1^-$ splitting, which results in  
$\epsilon(\mathrm{ABA})=2.4$ and $\epsilon(\mathrm{ABC})=2.7$.

\subsection{Tetralayer graphite}

The band structures near the K point of four types of stable and metastable 
tetralayer graphite films are shown in Fig.~\ref{band-4L}.
ABAC- and ACBC-stacked systems are obviously equivalent
at zero electric field and even at an uniform field along
c-axis they are equivalent when the reversed field is applied 
in either system. Therefore, the panel (k) may be thought of as 
the counterpart of panel (h) with reversed field.

The Bernal-type, ABAB-tetralayer system at the K point is equivalent
to a linear tetramer and two weakly bonded dimers configured
as is shown in panel (c). This clearly explains the distribution of 
Kohn-Sham eigenvalues at the K point in panel (a). 
The labels $4^\pm$ refer to 
the full bonding and antibonding states, 
and $4^{\pm*}$ are semi-bonding and antibonding states of
the linear tetramer. The external field lifts
the degeneracy of upper and lower dimers. However,
this does not leads to a band gap as show in panel (b)
at this level of the field strength. From the splitting
of weakly bonded dimeric states, the dielectric constant
in the system is evaluated to be $\epsilon(\mathrm{ABAB})=2.4$,
almost no change from the Bernal trilayer's value.

Similarly to the trilayer case, the rhombohedral-type
ABCA-tetralayer system develops a flat or rather concave 
band gap as an uniform
field comes in as shown in panel (e). 
The energy splitting of nearly non-bonding
states (labeled as 1) on the top and bottom 
layers is 0.263 eV, which corresponds to 
$\epsilon(\mathrm{ABCA})=3.0$. The induced narrow band gap
will be controlled using a variable field.

The ABAC- or ABCB-stacked system has a very narrow band gap 
structure as shown in panel (g), which is rather 
similar to the semi-metallic band of ABA-trilayer around the K point
at zero field, except for an overlaying dimeric band. 
When an external field is applied parallel to or anti-parallel 
to c-axis, the system becomes semi-metallic as plotted in
panels (h) and (k). The $3^0$ and $1^+$ states are always very 
close, nearly independent of the uniform field, since they feel 
nearly the same averaged external potential.
The dielectric constant is evaluated to be 
$\epsilon(\mathrm{ABAC})=2.6$ and $\epsilon(\mathrm{ABCB})=2.5$
at the present  finite field, where we notice
very small asymmetry due to the broken symmetry of the stacking sequence
along the c-axis.

\section{Conclusions}
\label{sec:Conclusion}

We have examined the stability of FLGs
and found that for any number of layers
Bernal FLGs are the most stable, but rhombohedral-type
tri- and tetra-layer graphite films are also predicted
to be possible structures, which are stabilized by 
relatively high energy barrier.
In the rhombohedral-type FLGs including
AB-bilayer, variable band gap will be induced
by the on-set of perpendicular electric field, and other
FLGs will stay semi-metallic.
(Quite recentry, a report on {\it ab initio} band structures of 
FLGs at zero field\cite{Latil06} 
came into our notice. We found that the present results at zero field is
in fine agreement with their results.)
Therefore, experimental challenges for the fabrication of
these non-Bernal FLGs would be very interesting.

\section*{Acknowledgments}
The authors thank the CPMD consortium for the code.

% \appendix

% \section{}
% \label{}


\begin{thebibliography}{00}

\bibitem{Novoselov-Sci04} K. S. Novoselov, A. K. Geim, S. V. Morozov, D. Jiang,
Y. Zhang, S. V. Dubonos, I. V. Grigorieva and A. A. Firsov, 
Science {\bf 306}, 666 (2004).

\bibitem{Zhang-nat05} Y. Zhang, Y.-W. Tan, H. L. Stormer and P. Kim, 
Nature {\bf 438}, 201 (2005).

\bibitem{Novoselov-NP06} K. S. Novoselov, E. McCann, S. V. Morozov, V. I. Falko, 
M. I. Katsnelson, U. Zeitler, D. Jiang, F. Schedin and A. K. Geim,
Nature Physics {\bf 2}, 177 (2006).

\bibitem{Zheng-Ando-PRB02}
Y. Zheng and T. Ando, 
Phys. Rev. {\bf B 65}, 245420 (2002).


\bibitem{Partoens-PRB06}
B. Partoens and F. M. Peeters, 
Phys. Rev. {\bf B 74}, 075404 (2006).

\bibitem{Lu-JPCM06} C. L. Lu, C. P. Chang, Y. C. Huang, J. M. Lu, 
C. C. Hwang and  M. F. Lin,
J. Phys.: Condens. Matter {\bf 18}, 5849 (2006).

\bibitem{Marinopoulos-PRB04}
A.~G. Marinopoulos, L. Reining, A. Rubio and V. Olevano, 
Phys. Rev. {\bf B 69}, 245419 (2004).


\bibitem{CPMD}
CPMD, version 3.9.2, Copyright IBM Corp 1990--2006, 
Copyright MPI f\"ur Festk\"orperforschung Stuttgart 1997--2001. 

\bibitem{Goedecker-PP-PRB96}
S. Goedecker,  M. Teter and J. Hutter, 
Phys. Rev. {\bf B 54} 1703 (1998).

\bibitem{Monkhorst-Pack-PRB76}
H. J. Monkhorst and J. D. Pack, Phys. Rev. {\bf B 13}, 5188 (1976).

\bibitem{Girifalco-PRB02}
L.~A. Girifalco and M. Hodak, Pack, Phys. Rev. {\bf B 65}, 125404 (2002).


\bibitem{Schabel-Gr-PRB92}
M.~C. Schabel and J.~L. Martins, 
Phys. Rev. {\bf B 46}, 7185 (1992) .

\bibitem{CharlierMG92}
J.-C. Charlier, J.-P. Michenaud and X. Gonze,
Phys. Rev. {\bf B 46}, 4531 (1992) .

\bibitem{Ohta-Sci06} T. Ohta, A. Bostwick, T. Seyller, H. Horn and E. Rotenberg,
Science {\bf 313}, 951 (2006).


\bibitem{Latil06} S. Latil and L. Henrard,
Phys. Rev. Lett. {\bf 97}, 036803 (2006).



% notes:

% \bibitem{label2}
% If there is a note, it should come last:
% \bibitem{label3} \note
% \end{subbibitems}

%%%\bibitem{}

\end{thebibliography}
\end{document}